\documentclass[superscriptaddress,amsmath,floatfix,aps,prc,twocolumn]{revtex4-2}

\usepackage{preamblestyle} 

\begin{document}

\title{Thermalization of Neutrinos in a Neutron Star Merger Simulation}

 \author{M.~G.~Alford \,\orcidlink{0000-0001-9675-7005}
}
 \email{alford@physics.wustl.edu}
 \affiliation{Department of Physics, Washington University in St.~Louis, St.~Louis, MO 63130, USA}
 
 \author{L.~Brodie\,\orcidlink{0000-0001-7708-2073
}}
 \email{b.liam@wustl.edu (corresponding author)}
 \affiliation{Department of Physics, Washington University in St.~Louis, St.~Louis, MO 63130, USA}

 \author{F.~Foucart\,\orcidlink{0000-0003-4617-4738
}}
\email{Francois.Foucart@unh.edu}
\affiliation{Department of Physics \& Astronomy, University of New Hampshire, 9 Library Way, Durham NH 03824, USA}
 
 \author{A.~Haber\,\orcidlink{0000-0002-5511-9565}}
 \email{a.haber@soton.ac.uk}
 \affiliation{Mathematical Sciences and STAG Research Centre, University of Southampton, Southampton SO17 1BJ, United Kingdom}
 \affiliation{Department of Physics, Washington University in St.~Louis, St.~Louis, MO 63130, USA}

\date{March 9, 2026} 

\begin{abstract}
We study the neutrino distributions that arise in a simulation of a neutron star merger that uses a Monte Carlo (MC) neutrino transport scheme. In a snapshot taken $1\,\ms$ after merger, we calculate relevant observables to test when neutrinos behave like a thermalized gas, and when a free-streaming picture is more appropriate.
We find that in hot, dense regions where neutrino–matter interactions are frequent, MC neutrino and antineutrino distributions are consistent with thermalized neutrinos. In moderately warm regions, where neither approximation is expected to hold, we find significant departures from the predictions of the thermalized-neutrino approximation, particularly for the (anti)neutrino average opacity and net rate of absorption per baryon, even when average energies appear approximately thermal. At lower temperatures, MC results approach the free-streaming limit. Our results demonstrate that energy-averaged agreement with thermalized-neutrino assumptions does not guarantee accurate weak interaction rates. Non-equilibrium aspects of the neutrino distribution are therefore crucial for neutrino-mediated microphysics such as composition evolution in the early post-merger phase.
\end{abstract}

\maketitle
\section{Introduction}
\label{sec:intro}
Neutron star mergers are unique astrophysical laboratories for probing the properties of ultra-dense matter at finite temperature. In these events, neutrinos play a central role in both the microphysical dynamics of the local hydrodynamical fluid and the macroscopic evolution of the merger remnant. Because neutrinos only couple to nuclear matter and themselves through the weak interaction, they are the particle species with the longest mean free path. Neutrinos therefore play a dominant role in out-of-equilibrium physics \cite{Rosswog:2003rv,Sekiguchi:2011zd,Baiotti:2016qnr,Sedrakian:2024uma, Benhar:2025tir}: they are the major source of radiative cooling \cite{Lattimer:1991ib,Yakovlev:2004iq}, mediate changes in composition through weak interactions \cite{Schmitt:2017efp,Alford:2020pld,Alford:2023gxq} (which can also lead to bulk viscous dissipation~\cite{Most:2022yhe,Espino:2023dei}),
and influence the dynamics of outflows, e.g., nucleosynthesis, that shape electromagnetic counterparts such as kilonovae \cite{Wanajo:2014wha,Palenzuela:2021gdo,Fujibayashi:2022ftg,Zappa:2022rpd,Sun:2022vri,Radice:2023zlw,Foucart:2024npn,Ricigliano:2024lwf,Qiu:2025ybw,Neuweiler:2025klw}. Accurate modeling of neutrino transport is therefore essential for connecting theoretical predictions to observations. 
 
As full Boltzmann neutrino transport is computationally demanding, many simulations of neutron star mergers, as well as phenomenological calculations of out-of-equilibrium dynamics, rely on assumptions about the neutrino transparency of matter in neutron stars. There are two commonly used assumptions. One is the free-streaming-neutrino approximation, where there are no neutrinos in initial states and no Pauli blocking due to neutrinos in final states. The other is the thermalized-neutrino approximation, where neutrinos are assumed to be in local thermal equilibrium with their surroundings, i.e., they are described by a Fermi–Dirac (FD) distribution defined by an ambient temperature and a neutrino chemical potential (also known as ``trapped'' in the literature \cite{Alford:2021lpp}). In the thermalized regime, neutrinos are often assumed to be in chemical equilibrium with the other constituents.

There are three main schemes for treating neutrinos in merger simulations: leakage, moment, and Monte Carlo (MC)~\cite{Foucart:2022bth}. The leakage scheme \cite{Ruffert:1996by,Rosswog:2003rv,Deaton:2013sla,Neilsen:2014hha} is the least computationally demanding approach and estimates local neutrino emission rates using the optical depth of nuclear matter, i.e., the opacity or inverse mean free path integrated along a characteristic path length. 
In general-relativistic merger simulations, the leakage scheme typically accounts only for a local rate of energy loss and composition change based on the local emissivity and diffusion rate of neutrinos through the remnant. It does not account for neutrinos emitted at one point and then absorbed or scattered at a different point in the simulation. 
Leakage schemes often interpolate between optically thin (free streaming) and optically thick (thermalized) limits in the calculation of weak interaction rates.

The moment scheme~\cite{Thorne:1981nvt,Shibata:2011kx} is more complex and computationally demanding than the leakage scheme; it allows for the emission, scattering, absorption, and non-local transport of neutrinos. It evolves angular moments of the neutrino energy distribution, e.g., the neutrino energy density and flux, by solving the angle-integrated Boltzmann equation. The less computationally demanding energy-integrated or ``grey'' moment scheme~\cite{Wanajo:2014,Foucart:2015vpa,Foucart:2016rxm,Radice:2021jtw} solves the energy-integrated Boltzmann equation. Energy-dependent moment schemes are being developed~\cite{Kuroda:2015bta,Cheong:2024buu}, but have not yet been implemented in general-relativistic neutron star merger simulations. Closure relations, which are required to truncate higher-order moments, depend on assumptions about the form of the neutrino distribution. Extracting energy-averaged quantities requires similar assumptions about the form of the distribution.
A common assumption is that neutrinos are in thermal equilibrium with their surroundings, with a temperature that is not necessarily the ambient temperature.

The MC scheme~\cite{Foucart:2021mcb} is a different neutrino transport scheme that directly samples the Boltzmann equation to evolve the energy- and angle-dependent neutrino distribution. For the number of packets used in current simulations~\cite{Foucart:2020qjb}, MC simulations have a computational cost comparable to that of two-moment simulations. This relatively low cost is
attainable because, even in the densest and hottest regions, MC simulations only need about 100 neutrino packets per cell to achieve a reasonably accurate picture of the long-term evolution of the fluid. In low-density regions, there may be fewer than one packet per cell on average.
MC simulations are thus sensitive to shot noise when attempting to reconstruct the neutrino distribution function in a small volume of phase space. They are, however, currently the only scheme that evolves the full neutrino distribution without assuming that it follows a specific functional form. 

In this paper, we present the first comparison between the predictions for physical observables from the neutrino distributions
yielded by MC simulation data and predictions from the free-streaming-neutrino and thermalized-neutrino approximations. In Sec.~\ref{sec: methods}, we describe the neutron star merger simulation and its MC treatment of neutrinos and define the quantities we calculate to compare the different neutrino population predictions. In Sec.~\ref{sec: compare_mc_fd_hot}, we compare the neutrino and antineutrino average energy, average absorption opacity, and net rate of absorption per baryon for some of the hottest fluid cells in the simulation data ($T>60\,\MeV$). We do this to probe the MC neutrino distributions in a regime where the thermalized-neutrino approximation should be valid. In Sec.~\ref{sec: compare_mc_fd_warm}, we compare the same quantities but for more moderate temperatures where neither the thermalized-neutrino approximation nor the free-streaming-neutrino approximation are likely to be good assumptions. 

We use natural units with $\hbar = c = k_B = 1$.

\section{Methods}
\label{sec: methods}

\subsection{Simulation Details}
\label{sec: simulation_details}

We obtain information about the distribution function of neutrinos in a binary neutron star merger system from a simulation performed with the Spectral Einstein Code (SpEC)~\cite{Duez:2008rb}, specifically a snapshot of simulation ``MC-HR'' (MC-High Resolution) of Ref.~\cite{Foucart:2024npn}, taken $1\,\ms$ after merger. The local merger time is defined as the time when the maximum baryon density on the grid reaches 1.03 times the initial maximum density. The simulation we consider evolves neutron stars with masses $(1.4\,\mathrm{M}_\odot,1.3\,\mathrm{M}_\odot)$, using the SFHo equation of state~\cite{Steiner:2012rk}. SpEC evolves Einstein's equations of General Relativity in the generalized harmonic formalism~\cite{Lindblom:2005qh}, together with the fluid equations in the Valencia formalism~\cite{Banyuls:1997zz}, and Boltzmann's equation for neutrino transport using an energy-dependent MC transport algorithm~\cite{Foucart:2021mcb}. Matter within the neutron star is described by its temperature, baryon density, electron fraction, and velocity. Magnetic fields, muons, and neutrino flavor oscillations are ignored in the simulation. 

In the SpEC simulations, neutrino-matter interaction rates are computed through tabulated values of an energy-dependent emissivity, absorption opacity, and elastic scattering opacity, all generated using the \texttt{NuLib} library~\cite{nulib} with the SFHo equation of state. 
The simulation evolves only three distinct neutrino species: $\nu_e$, $\bar \nu_e$, and a $\nu_x$ group that
accounts for all other species ($\nu_\mu,\bar\nu_\mu,\nu_\tau,\bar\nu_\tau)$.
The tabulated reactions in the simulation include capture of electron neutrinos and antineutrinos on neutrons and protons, as well as their inverse reactions,
\begin{equation}
\begin{aligned}
    n+\nu_e &\leftrightarrow p+e^- \, ,\\
    p+\nubar_e&\leftrightarrow n + e^+ \, .
\end{aligned}
\label{eq: xp_change_processes}
\end{equation}
They also include scattering of all flavors of neutrinos and antineutrinos on neutrons, protons, alpha particles, and heavy nuclei:
\begin{equation}
\begin{aligned}
    B + \nu_{e,x} &\leftrightarrow B + \nu_{e,x} \, ,\\
    B + \nubar_{e,x} &\leftrightarrow B + \nubar_{e,x} \, .
\end{aligned}
\label{eq: scattering_processes}
\end{equation}
The scattering reactions assume elastic scattering, i.e., they do not change the (anti)neutrino energy (by which we mean neutrinos or antineutrinos). For heavy lepton (but not electron) neutrinos and antineutrinos, the simulation also includes electron-positron annihilation and nucleon-nucleon bremsstrahlung, with rates computed assuming the neutrinos and antineutrinos are thermally equilibrated
\begin{equation}
\begin{aligned}
    e^+ + e^- &\leftrightarrow \nu_{x} + \nubar_{x} \, , \\
    N+N &\leftrightarrow N+N + \nu_{x} + \nubar_{x} \, .
\end{aligned}
\label{eq: heavy_nu_nubar_pair_processes}
\end{equation}
Within \texttt{NuLib}, we turn on flags for weak magnetism corrections, ion-ion correlations, heavy scattering form factors, electron polarization correction, and transport opacities. 

The results presented in Sec.~\ref{sec:results} use a snapshot taken very shortly after the merger to observe the behavior of the neutrinos in moderately warm, dense regions, which are present in the system during the first few milliseconds post-merger, before the post-merger remnant becomes more uniformly hot.

\begin{figure}[h]
\includegraphics[width=\columnwidth]{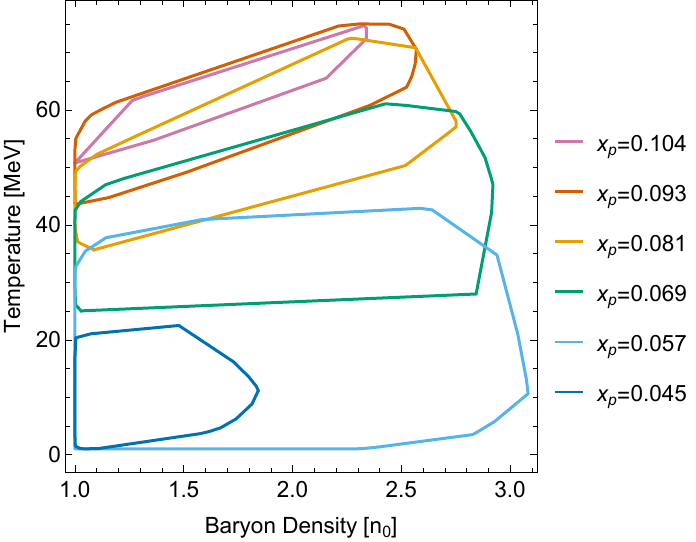}
\caption{
Each contour encloses the range of temperature and baryon number density of fluid cells with a proton fraction within $\pm\,5\%$ of the specified \texttt{NuLib} proton fractions $x_p$, in our 1\,\ms\ time slice.
}
\label{fig: fluid_xp_contours}
\end{figure}
 
The fluid parameter space from the $1\,\text{ms}$ post-merger time slice is shown in Fig.~\ref{fig: fluid_xp_contours},
in which each contour encloses the range of densities and temperatures found in fluid cells with a given proton fraction (by charge neutrality, this is the same as the electron fraction in $npe$ matter), 
selected from the \texttt{NuLib} grid points with $\pm\,5\%$ tolerance. 
For example, we can see immediately that among fluid cells with baryon density $n_B = 2\,\nsat$ and temperature $T = 30\,\MeV$,
if we look for those whose proton fraction falls within $\pm 5\%$ of a \texttt{NuLib} grid point, we will only find the $x_p = 0.057$ and $x_p = 0.069$ grid points represented. Above we have defined baryon saturation density $\nsat=0.16\,\text{fm}^{-3}$ and proton fraction $x_p=n_p/n_B$, where $n_p$ is the net proton minus antiproton number density. For reference, the neutrinoless equilibrium proton fraction for SFHo at zero temperature and the range of baryon densities shown is $x_p \approx 0.06$. 

\begin{figure}
    \centering
    \includegraphics[width=\linewidth]{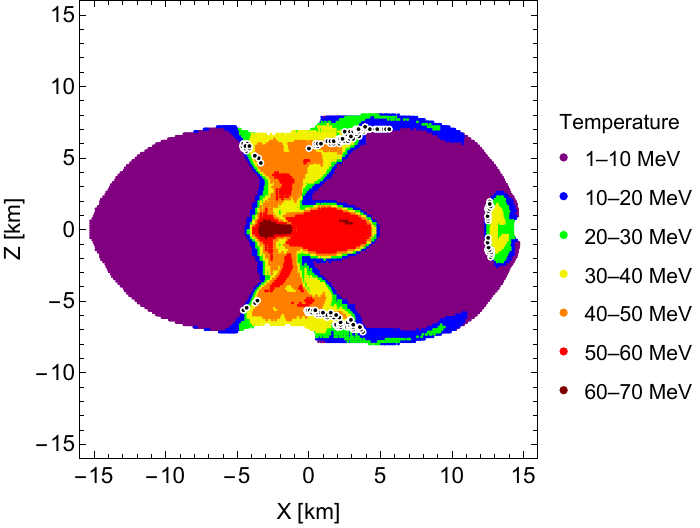}

    \vspace{0.4em}
    \includegraphics[width=\linewidth]{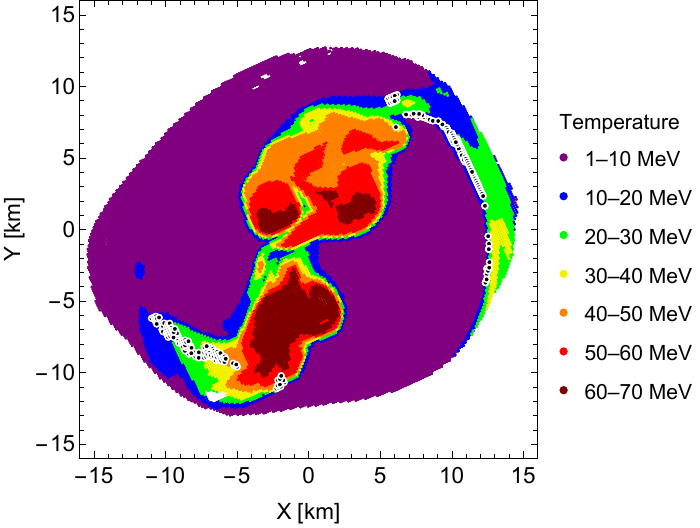}
    \caption{Top: Side cross-sectional view of the thermal profile for the merger simulation $1\,\ms$ after merger showing all fluid cells with $n_B\geq1\,\nsat$, where the rotation axis is in the $z$-direction. The black dots show the location of a subset of the fluid cells we analyze in Sec.~\ref{sec: compare_mc_fd_warm}.\\
    Bottom: Same as the top panel but showing an equatorial cross-sectional view.
    }
    \label{fig: simulation_profile}
\end{figure}

We show cross-sectional thermal profiles of the neutron star merger simulation $1\,\ms$ after merger in Fig.~\ref{fig: simulation_profile}, where the rotation axis is in the $z$-direction. In the top panel, we show a side cross-sectional view ($y=0$), and in the bottom panel, we show an equatorial cross-sectional view ($z=0$). All of the fluid cells shown have a baryon density greater than nuclear saturation density and have a width of about $200\,\text{m}$. The individual stars can still be distinguished as the two cold regions mirrored about the $x$-axis. We show the location of a subset of the individual fluid cells we analyze in Sec.~\ref{sec: compare_mc_fd_warm} ($T=12.3\,\MeV$, $T=17.9\,\MeV$, and $T=33.4\,\MeV$ with $n_B=1.78\,\nsat$ and $x_p=0.057$) as black dots. The location of these fluid cells is between the cold regions, where the free-streaming-neutrino approximation is typically valid, and the hot interface region, where the thermalized-neutrino approximation is typically valid. 

\subsection{Fluid Cell Properties}
\label{sec:cell-properties}

To make meaningful comparisons, we need to study subpopulations of cells grouped by common physical properties. The thermodynamic properties that characterize a fluid cell are its temperature $T$, baryon density $n_B$, proton fraction $x_p = n_p/n_B$, electron-type neutrino fraction $x_\nu = n_\nu/n_B$, and electron-type antineutrino fraction $x_\nubar = n_\nubar/n_B$. 
References to (anti)neutrino number densities and lepton fractions throughout this paper are to the electron type.
From these quantities, we can compute the lepton fraction
\begin{equation}
    \label{eq: lepton_fraction}
    x_L \equiv \dfrac{n_{e^-} + n_\nu - n_{\bar{\nu}}}{n_B} \ ,
\end{equation}
where $n_{e^-}$ is the net electron (electron minus positron) number density. Note that $n_\nu$ ($n_{\nubar}$) is the neutrino (antineutrino) number density, not the net neutrino number density.

One of our goals is to compare the MC simulation data, where the neutrino population is not necessarily in thermal equilibrium, with the predictions of the thermalized-neutrino approximation in which the (anti)neutrinos are assumed to follow a Fermi--Dirac distribution. Specifically, we compare the MC results for fluid cells in the simulation to the predictions for a cell that has the same temperature, baryon density, and proton fraction, but where the (anti)neutrino spectra are Fermi--Dirac distributions 
$f^{\text{FD}}(k,\mu,T) = 1/(1+\exp((E(k)-\mu)/T))$
whose temperatures are the same as those of the fluid cell and whose chemical potentials $\mu_\nu$ and $\mu_\nubar$ are obtained by solving
\begin{equation}
\ba{rcl}
\dsp\int_0^\infty \!\!\frac{d^3k}{(2\pi)^3}\,f^{\text{FD}}(k,\mu_\nu,T) &=& n^\text{MC}_\nu \, ,\\[3ex]
\dsp\int_0^\infty \!\!\frac{d^3k}{(2\pi)^3}\,f^{\text{FD}}(k,\mu_\nubar,T) &=& n^\text{MC}_\nubar \, ,
\ea
\label{eq:munu-def}
 \end{equation}
such that they give the same (anti)neutrino number densities as in the MC fluid cell.
We do not assume that $\mu_\nubar=-\mu_\nu$ because in \texttt{NuLib} the reactions that scatter (anti)neutrinos do not change their energies, so these reactions do not contribute to thermalization of (anti)neutrinos or chemical equilibration to $\mu_\nubar=-\mu_\nu$.
In particular, the merger simulation we consider also does not include electron neutrino and antineutrino pair creation and annihilation (see Eq.~\eqref{eq: heavy_nu_nubar_pair_processes} for the case with heavier lepton neutrinos and antineutrinos), so there is no way for the thermalization of neutrinos to influence antineutrinos and vice versa. The only way to achieve $\mu_\nu = -\mu_\nubar$ in this simulation is through equilibration of both reactions in Eq.~\eqref{eq: xp_change_processes}.

Where it is physically meaningful (i.e., for the net rate of (anti)neutrino absorption per baryons, see Sec.~\ref{sec:proton-fraction}), we will also compare the MC results with the predictions of the free-streaming approximation, i.e., no (anti)neutrinos in initial states and no Pauli blocking of final states. 

\subsection{Energy-Averaged Observables}
\label{sec: energy-averaged}
One metric for comparing the MC results for a given population of fluid cells with the thermalized-neutrino approximation is the average (anti)neutrino energy and opacity in each cell.

The average energy of the (anti)neutrinos in a single fluid cell $c$ is 
\begin{equation}
     \langle E \rangle_c  \equiv \frac{\sum_{p\in c} E_p\, N_p}{\sum_{p\in c} N_p} \, ,
    \label{eq: avg_e}
\end{equation}
where $p$ is summed over (anti)neutrino MC packets in a cell and $N_p$ is the number of (anti)neutrinos in the $p$-th packet, in cell $c$. 
All (anti)neutrinos in an MC packet have the same energy $E_p$, and every (anti)neutrino packet in the simulation is created with the same fluid-frame energy $E_{\,\text{packet}} = E_p N_p$, which may then evolve via relativistic effects (e.g.,~gravitational redshift).
Ignoring relativistic effects (which are likely not large except at points where the fluid is moving close to the speed of light), we can write the average energy of (anti)neutrinos in a single fluid cell as
\begin{equation}
     \langle E \rangle_c  = \frac{P_{c}\,E_{\,\text{packet}}}{\sum_{p\in c} N_p} \, ,
    \label{eq: avg_e_using_e_tot}
\end{equation}
where $P_{c} = \sum_{p\in c}1$ is the total number of packets in cell $c$.

\begin{table*}[htbp]
\centering
\begin{tabular}{|c|*{16}{c|}}
\hline
$E_{i}$ [MeV] & 2.0 & 6.0 & 10.5 & 16.3 & 23.7 & 33.0 & 44.8 & 59.9 & 79.0 & 103.4 & 134.3 & 173.6 & 223.6 & 287.1 & 367.8 & 470.4 \\
\hline
$\Delta E_{i}$ [MeV] & 4.0 & 4.0 & 5.1 & 6.5 & 8.2 & 10.4 & 13.3 & 16.9 & 21.4 & 27.2 & 34.6 & 44.0 & 55.9 & 71.1 & 90.4 & 114.8 \\
\hline
\end{tabular}
\caption{(Anti)neutrino energy bin centers $E_{i}$ and widths $\Delta E_{i}$, taken from the neutrino reaction library \texttt{NuLib}~\cite{nulib}.}
\label{tab: energy_bins}
\end{table*}

The average (anti)neutrino absorption opacity is defined in a similar way to the average energy, except the \texttt{NuLib} opacity table the simulation uses (and we also use) \cite{nulib} is tabulated with a finite grid in energy, so we use the same (anti)neutrino energy bins (see Table~\ref{tab: energy_bins})
\begin{equation}
    \langle \kappa \rangle_c \equiv \frac{\sum_i \kappa(E_i)\, n_c(E_i) }{\sum_i n_c(E_i)},
    \label{eq: avg_opacity}
\end{equation}
where $i$ is summed over (anti)neutrino energy bins and $n_c(E_i)$ is the number density of (anti)neutrinos in the $i$-th energy bin, in cell $c$.
Note that \texttt{NuLib} directly provides the stimulated absorption opacity. We use the bare absorption opacity, see Sec.~\ref{sec:proton-fraction}.
The (anti)neutrino absorption opacity $\kappa(T, n_B, x_p, E)$ depends on the fluid cell parameters as well as the (anti)neutrino energy. 

\subsection{Rate of Neutrino Absorption and Emission}
\label{sec:proton-fraction}
Another quantity that we can calculate for each MC fluid cell, and compare with the thermalized-neutrino approximation as well as the free-streaming-neutrino approximation, is the net rate of (anti)neutrino absorption per baryon $\gamma_{\nu,\nubar}$.
We consider weak interactions that can change the net neutrino fraction (and therefore the proton fraction) for matter consisting of neutrons, protons, and electrons, see Eq.~\eqref{eq: xp_change_processes}.
The net rate per volume of neutrino absorption is denoted \mbox{$\Gamma^{n \nu \leftrightarrow p e^-}$} and the net rate per volume of antineutrino absorption \mbox{$\Gamma^{p \nubar \leftrightarrow n e^+}$}, where we define \mbox{$\Gamma^{a\leftrightarrow b} \equiv \Gamma^{a\rightarrow b}- \Gamma^{a\leftarrow b}$}. To get an inverse time scale that can be compared to the dynamical time scale of a neutron star merger, we compute the net (anti)neutrino absorption rate per baryon 
\begin{align}
    \gamma_{\nu} &\equiv \frac{\Gamma^{n \nu \leftrightarrow p e^-}}{n_B}\, 
    \label{eq:gamma-nu} \, ,\\
    \gamma_{\nubar} &\equiv \frac{\Gamma^{p \nubar \leftrightarrow n e^+}}{n_B}\, .
   \label{eq:gamma-nubar}
\end{align}

To find the net rate per unit volume $\Gamma$ of each reaction, we convolve the (anti)neutrino absorption opacity and emissivity with an (anti)neutrino distribution. That distribution can be: 
\begin{itemize}
    \item[(a)] Obtained from the MC simulation data;
    \item[(b)] Derived by assuming (anti)neutrinos are thermally equilibrated as described in Sec.~\ref{sec:cell-properties};
    \item[(c)] Derived by assuming free-streaming neutrinos, i.e., no (anti)neutrinos in initial states and no Pauli blocking of final states. 
\end{itemize}

The net rate per volume of antineutrino absorption is 
\begin{equation}
    \Gamma^{p  \nubar \leftrightarrow n e^+} {=} \int \!\frac{d^3\mathbf{k}_{\nubar}}{(2\pi)^3} \kappa_\nubar(\mathbf{k}_{\nubar}) f_{\nubar}(\mathbf{k}_{\nubar}) - \eta_\nubar (\mathbf{k}_{\nubar}) \big(1{-}f_{\nubar}(\mathbf{k}_{\nubar})\big) \,,
    \label{eq: gamma_net_nubar}
\end{equation}
where $\mathbf{k}_{\nubar}$ is the three-momentum, $\eta_\nubar$ is the emissivity, $\kappa_\nubar$ is the opacity, $f_{\nubar}$ is the occupation number, and $(1-f_{\nubar})$ is the final state Pauli blocking factor. 

Since \texttt{NuLib} uses energy bins for neutrinos (see Table~\ref{tab: energy_bins}), we need to convert Eq.~\eqref{eq: gamma_net_nubar} into a sum over energy bins. We assume the antineutrinos are distributed isotropically in momentum space and that the antineutrinos are massless, i.e., $E_{\nubar}=|\mathbf{k}_{\nubar}|$, the same assumptions that are used in the merger simulation we consider. 
We define the (anti)neutrino Pauli blocking factor (including the integration measure)
\begin{equation}
    B_{\nu,\nubar}(E_i) \equiv \frac{\Delta E_i E_i^2}{2\pi^2} - n_{\nu,\nubar}(E_i)\, ,
\end{equation}
where $0 \leq B_{\nu,\nubar}(E_i) \leq \Delta E_i E_i^2/2\pi^2$, because $n_{\nu,\nubar}$ cannot exceed the available phase space. 
The net rate per volume of antineutrino absorption is now
\begin{align}
     \Gamma^{p \nubar \leftrightarrow n e^+}
     \approx \sum_{i=1}^{N_{\text{bins}}} \kappa_\nubar(E_i)n_{\bar{\nu}}(E_i) - \eta_\nubar (E_i) B_\nubar(E_i)\,,
     \label{eq: summed_net_anu}
\end{align}
which is exact in the limit ${\Delta E_i \rightarrow 0}$ and when summing over energies extending to infinity.
Similarly, the net rate per volume of neutrino absorption is 
\begin{align}
\label{eq: summed_ec}
     \Gamma^{n \nu \leftrightarrow p e^-}
     \approx \sum_{i=1}^{N_{\text{bins}}} \kappa_\nu(E_i)n_{\nu}(E_i) - \eta_\nu(E_i) B_\nu(E_i) \, .
\end{align}

A relationship between the antineutrino emissivity and opacity can be found by considering Eq.~\eqref{eq: gamma_net_nubar} in equilibrium, i.e., when the integrand is zero for all values of $\mathbf{k_{\nubar}}$. We can therefore write
\begin{align}    
\label{eq: opacity_to_emissivity}
    \eta_\nubar (\mathbf{k}_{\nubar}) &= \kappa_\nubar(\mathbf{k}_{\nubar})\frac{f^{\text{FD}}_{\nubar}(\mathbf{k}_{\nubar})}{1-f^{\text{FD}}_{\nubar}(\mathbf{k}_{\nubar})} \\
    &= \kappa_\nubar (\mathbf{k}_{\nubar})\, e^{-(|\mathbf{k}_{\nubar}|-\delta\mu)/T} \nonumber,
\end{align}
where $\delta\mu=\mu_n-\mu_p-\mu_e$. In equilibrium, $\delta\mu = \mu_{\bar{\nu}}$, but notice that Eq.~\eqref{eq: opacity_to_emissivity} holds out of equilibrium as well \cite{Martinez-Pinedo:2012eaj}. 

For $n + \nu \leftrightarrow p + e^-$, the relationship between the neutrino emissivity and opacity is the same as in Eq.~\eqref{eq: opacity_to_emissivity} but with $\delta\mu = -(\mu_n-\mu_p-\mu_e)$ such that $\delta\mu=\mu_\nu$ in equilibrium. With Eq.~\eqref{eq: opacity_to_emissivity}, it is possible to write the net rate of (anti)neutrino absorption per volume using only information about the (anti)neutrino distribution and opacity.

To be consistent with the simulation, we use opacities from \texttt{NuLib}~\cite{nulib}. 
\texttt{NuLib} directly provides the stimulated absorption opacity $\kappa^*$, which is a useful quantity for transport calculations (see, e.g., Appendix B in Ref.~\cite{Thompson2003}), but is not the inverse mean free path of the (anti)neutrinos. The stimulated absorption opacity is related to the bare absorption opacity $\kappa$ (the inverse mean free path) by
\begin{equation}
\label{eq: stimulated_opacity}
    \kappa_{\nu,\nubar}^*(E_{\nu,\nubar}) = \frac{\kappa_{\nu,\nubar}(E_{\nu,\nubar})}{1-f^{\text{FD}}(E_{\nu,\nubar},\mu_{\nu,\nubar}^{\text{eq}},T)}, 
\end{equation}
where $\mu_{\nu,\nubar}^{\text{eq}}$ is the (anti)neutrino chemical potential in chemical equilibrium. For our calculations, we use Eq.~\eqref{eq: stimulated_opacity} to get the bare absorption opacity. We then use Eq.~\eqref{eq: opacity_to_emissivity} to get the emissivity.

\section{Results}
\label{sec:results}
We now compare properties of the electron-type neutrino and antineutrino distributions obtained from the MC simulation data described in Sec.~\ref{sec: simulation_details}
with the predictions made by two commonly used approximations: (a)~thermally-equilibrated neutrinos described by a Fermi--Dirac distribution; (b)~free-streaming neutrinos. We use data from the MC simulation $1\,\ms$ after merger, and examine differences in the (anti)neutrino average energy, average absorption opacity, and rate of absorption and emission.

\subsection{Neutrino Thermalization}
\label{sec: thermalization}

\begin{figure}[h!]
\centering

\includegraphics[width=\linewidth]{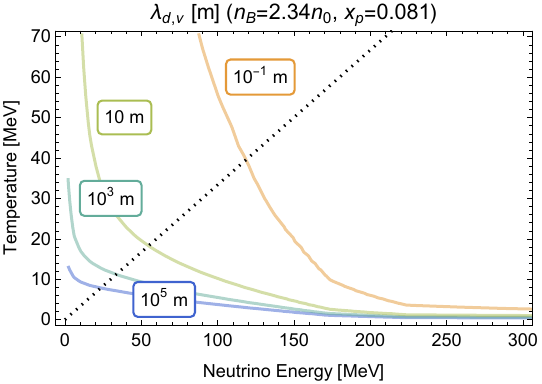}

\vspace{0.5em}

\includegraphics[width=\linewidth]{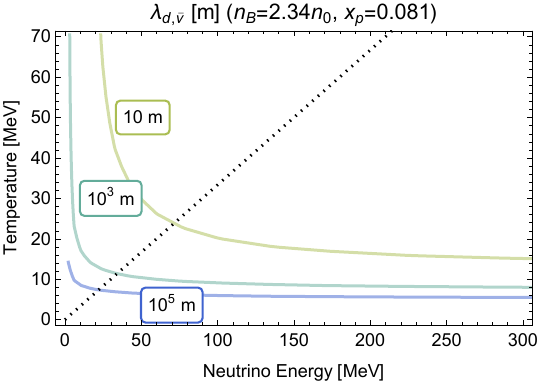}

\caption{The (anti)neutrino mean net displacement between thermalizing collisions (Eq.~\eqref{eq:lambda-d}) (in meters) for $n_B=2.34\,\nsat$ and $x_p=0.081$. The black dotted lines show the average energy $\langle E_{\nu,\nubar} \rangle=3 T$ of nondegenerate thermalized (anti)neutrinos. The qualitative behavior of both panels does not change if the proton fraction or the baryon density take the other values considered in this paper, i.e., $x_p=0.057$ and $n_B=1.78\,\nsat$.
}
\label{fig:mfp-diffusion}
\end{figure}

In a real-world merger (or a very accurate MC simulation), we would expect the (anti)neutrinos to be thermalized to the same temperature as the surrounding nuclear matter when the
mean net displacement (see Eq.~\eqref{eq:lambda-d} below) between thermalizing collisions is much shorter than the distance scale over which the temperature varies significantly. Additionally,  the mean time between thermalizing collisions has to be much shorter than the time scale over which the temperature varies significantly. 

To estimate the temperature variation distance, we show in Fig.~\ref{fig: simulation_profile} the thermal profile for a cross-sectional view of the merger $1\,\ms$ after merger. We show as black dots the location of the fluid cells with temperatures $T=12.3\,\MeV$, $T=17.9\,\MeV$, and $T=33.4\,\MeV$ that we include in our analysis in Sec.~\ref{sec: compare_mc_fd_warm}. These fluid cells are located in regions where the temperature gradient is about $10\,\MeV/\km$, so the characteristic distance for significant temperature variation can be estimated as $T/(dT/dx)$, which is in the km range.
We therefore expect that when the (anti)neutrino mean net displacement between thermalizing collisions is much less than 1\,\km\ they should be thermalized.

The mean net displacement between thermalizing collisions is 
\begin{equation}
 \la_d \equiv \min(\la_a, \sqrt{\la_a\la_s}) \ ,
 \label{eq:lambda-d}
\end{equation}
where
$\la_a$ is the absorption mean free path and $\la_s$ is the scattering mean free path. We have used the fact that there are, on average, $N=\la_a/\la_s$ elastic scattering events between emission and absorption, so we have a mean net displacement $\la_s\sqrt{N}$ (see Eq.~(25) in Ref.~\cite{Ardevol-Pulpillo:2018btx}). In Fig.~\ref{fig:mfp-diffusion} we show the mean net displacement computed using \texttt{NuLib} for neutrinos (top panel) and antineutrinos (bottom panel) as a function of (anti)neutrino energy and temperature, for nuclear matter at a typical density $n_B=2.34\,\nsat$ and proton fraction $x_p=0.081$. The dotted lines show the average (anti)neutrino energy $\langle E_{\nu,\nubar} \rangle= 3T$ in a thermal distribution of temperature $T$ and negligible chemical potential. This should be a reasonable guide in the nondegenerate case $\mu_{\nu,\nubar} \ll T$. This is true for the neutrinos at $T=62.5\,\MeV$ (the temperature we analyze in Sec.~\ref{sec: compare_mc_fd_hot}), where $\mu_{\nu}\lesssim -80\,\MeV$. At $T=62.5\,\MeV$, the typical antineutrino chemical potential is $\mu_\nubar \approx 110\,\MeV$, so the $\langle E_\nubar \rangle = 3T$ line is a lower bound on the average antineutrino energy.

As noted above, we expect neutrinos to be thermalized when their mean net displacement between thermalizing collisions is much less than 1\,\km. From Fig.~\ref{fig:mfp-diffusion} we see that it is difficult to formulate a general criterion for a thermalization temperature because the mean net displacement depends strongly on the neutrino energy. However, focusing on the (anti)neutrinos of average energy (black dotted line) and higher, we conclude that thermalization is expected at $T\gtrsim 20\,\MeV$ for neutrinos and antineutrinos.
We now proceed in Sec.~\ref{sec: compare_mc_fd_hot} with a discussion of the hottest fluid cells in the simulation, which meet the conditions for (anti)neutrino thermalization. In Sec.~\ref{sec: compare_mc_fd_warm} we discuss fluid cells whose temperature is closer to and even below the thermalization temperature estimated above.

\subsection{High Temperature}
\label{sec: compare_mc_fd_hot}

\begin{figure}[h!]
\includegraphics[width=\linewidth]{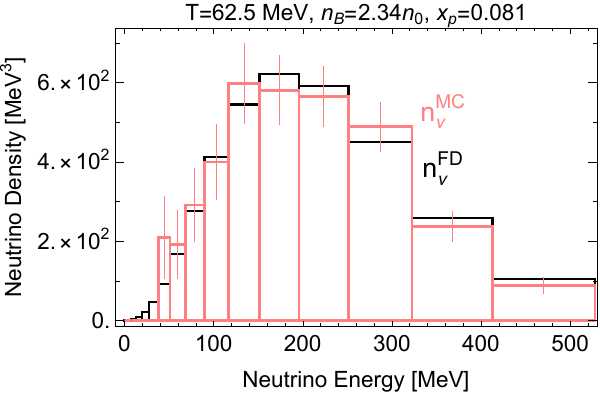}

\vspace{0.25em}

\includegraphics[width=\linewidth]{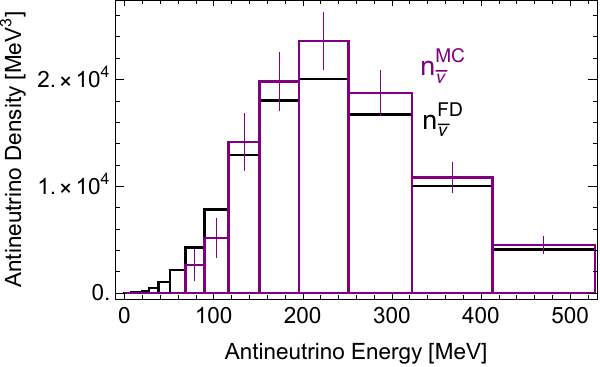}

\caption{Top: Neutrino number density as a function of energy. The energy-binned Monte Carlo (MC) distribution in pink is from a representative fluid cell with parameters $T = 62.5\,\MeV$, $n_B = 2.34\,\nsat$, $x_p = 0.081$, $x_L = 0.047$ ($x_\nu=0.001$, $x_\nubar=0.035$). The Fermi--Dirac (FD) distribution, where $\mu_\nu$ is fixed by the fluid cell's neutrino number density, is in black.\\
Bottom: Antineutrino number density as a function of energy in purple from the same fluid cell as the top panel, where $\mu_\nubar$ is fixed by the fluid cell's antineutrino number density.}
\label{fig: high_temp_nu_dist}
\end{figure}

\begin{figure*}[htbp]
\centering
\includegraphics[width=0.49\linewidth]{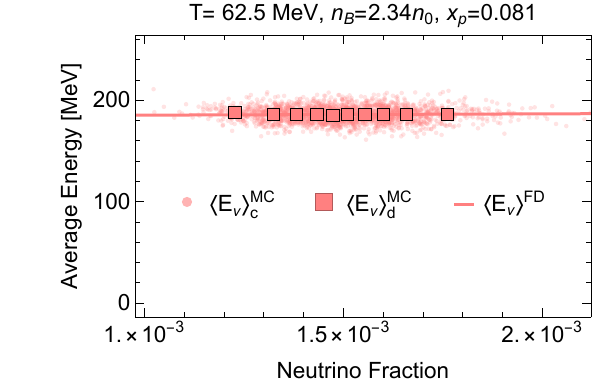}
\includegraphics[width=0.49\linewidth]{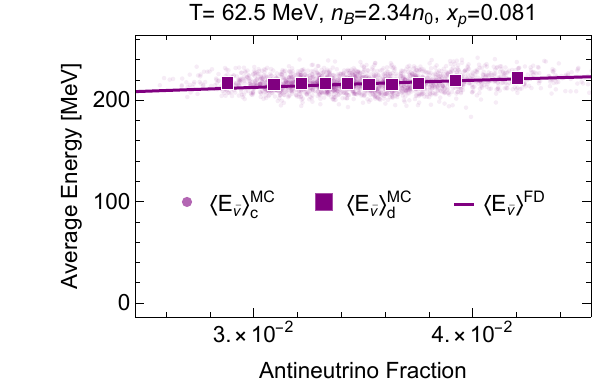}

\vspace{1em}


\includegraphics[width=0.49\linewidth]{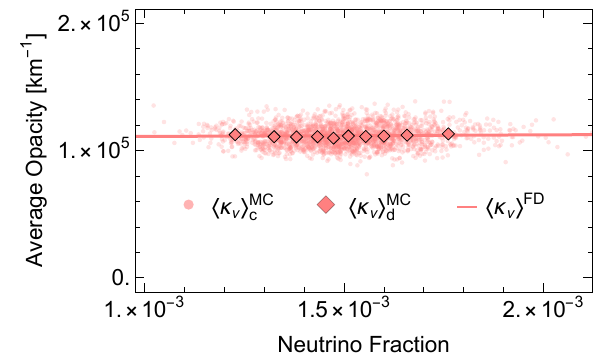}
\includegraphics[width=0.49\linewidth]{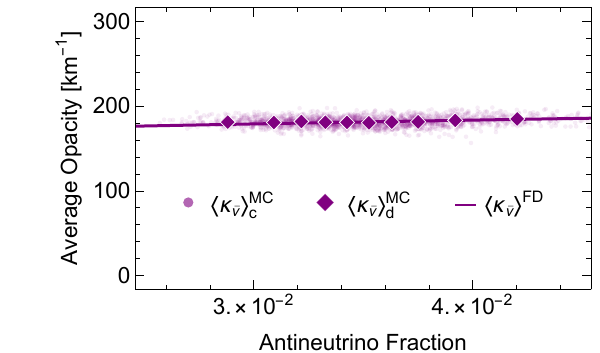}

\vspace{1em}

\includegraphics[width=0.49\linewidth]{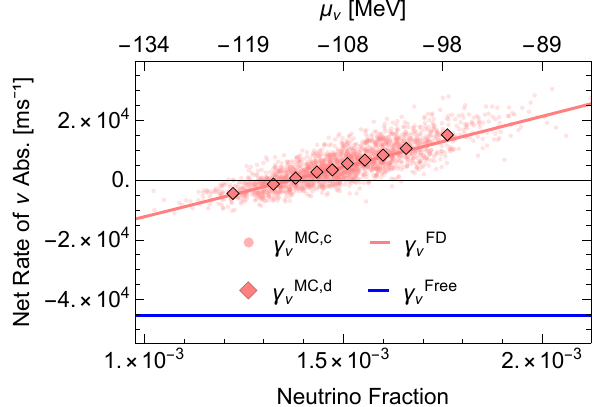}
\includegraphics[width=0.49\linewidth]{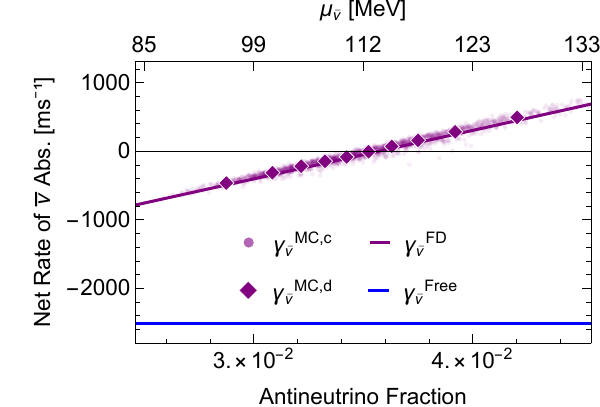}

\caption{Properties of hot ($T=62.5\,\MeV$) fluid cells with baryon density $n_B=2.34\,\nsat$ and proton fraction $x_p=0.081$, comparing Monte Carlo (MC) simulation data with predictions from thermalized-neutrino (FD) and free-streaming-neutrino (Free) approximations.\\
(a) Top row: Average neutrino (left panel, pink) and antineutrino (right panel, purple) energy as a function of the (anti)neutrino fraction, showing individual MC fluid cells (pale dots), averages over cells in (anti)neutrino fraction deciles,
and the thermalized-neutrino approximation (solid pink/purple lines).\\
(b) Middle row: Average (anti)neutrino opacity, following
the same conventions as the top row, except the diamonds correspond to decile medians.\\
(c) Bottom row: Net rate of (anti)neutrino absorption per baryon as a function of (anti)neutrino fraction and the corresponding chemical potentials in the thermalized-neutrino approximation, following
the same conventions as the top row, except the diamonds correspond to decile medians. Blue lines are the predictions of the free-streaming-neutrino approximation.
}
\label{fig: high_temp_grid}
\end{figure*}

We will now study the neutrinos and antineutrinos in the fluid cells in the \mbox{$T=62.5\,\MeV$} bin: this is the second highest \texttt{NuLib} temperature grid point that arises in the MC simulation time slice we consider ($t=1\,\ms$ after merger). Note that the highest temperature grid point has too few fluid cells to yield useful results. At \mbox{$T=62.5\,\MeV$}, as discussed in Sec.~\ref{sec: thermalization}, we expect the neutrinos and antineutrinos in the MC simulation data to be thermalized. 

In Fig.~\ref{fig: high_temp_nu_dist} (described in Sec.~\ref{sec:hot-neutrinos}), we show an example MC neutrino and antineutrino distribution, alongside the prediction of the thermalized-neutrino approximation.
Then, in Fig.~\ref{fig: high_temp_grid} (described in Secs.~\ref{sec: average_nu_e_hot_results}, \ref{sec:hot-nu-avg-opacity}, and \ref{sec:hot-neutrino-absorption}) we check the degree of thermalization by calculating several observables that are relevant to merger dynamics, and compare the MC simulation data with the predictions of the thermalized-neutrino and free-streaming-neutrino approximations. For those comparisons, we use data for the subset of MC simulation fluid cells whose fluid parameters (within a $\pm\, 5\%$ tolerance) are $T=62.5\,\MeV$, $n_B=2.34\,\nsat$, and $x_p=0.081$. This density and proton fraction were chosen because of the large number of fluid cells in that part of the parameter space and because they lie on the \texttt{NuLib} grid we use to evaluate the opacities and rates. 

\subsubsection{Hot Neutrino Energy Spectrum}
\label{sec:hot-neutrinos}

To give a sense of the MC (anti)neutrino distributions underlying our comparisons, we choose a representative fluid cell and show in Fig.~\ref{fig: high_temp_nu_dist} its neutrino and antineutrino energy spectrum (pink/purple histograms),
alongside the predictions of the thermalized-neutrino
approximation (black histograms), i.e., a Fermi--Dirac distribution, discretized into the same \texttt{NuLib} bins, see Table~\ref{tab: energy_bins}. We see that the shape of the MC neutrino and antineutrino spectrum is similar to the Fermi--Dirac distribution, as expected for thermalized neutrinos. The error bars are $n(E_i)/\sqrt{P_i}$, where $P_i$ is the number of MC (anti)neutrino packets in an energy bin $E_i$.

\subsubsection{Average Neutrino Energy}
\label{sec: average_nu_e_hot_results}
In the top row of Fig.~\ref{fig: high_temp_grid}, we show the average neutrino (left panel, pink) and antineutrino (right panel, purple) energies as a function of the (anti)neutrino 
fraction $x_{\nu,\nubar}$. The pale dots are averages within individual MC fluid cells $\langle E \rangle_c^\text{MC}$ computed using Eq.~\eqref{eq: avg_e}.
The solid pink/purple lines are predictions of the thermalized-neutrino approximation for the average (anti)neutrino energy, i.e., from Fermi--Dirac distributions whose $\mu_\nu$ or $\mu_\nubar$ are determined by the (anti)neutrino fraction in the fluid cell, see Eq.~\eqref{eq:munu-def}. 

We also plot decile averages of the MC data for neutrinos (pink squares) and antineutrinos (purple squares) with statistical error bars that are smaller than the plot markers. Note that we divide the fluid cells into deciles based on their (anti)neutrino fractions. In each decile, we average $\langle E \rangle^{\text{MC}}_c$ over the fluid cells in that decile, weighting each cell by its total number density of (anti)neutrinos 
\begin{equation}
  n_c = \sum_i n_c(E_i)\ .
\end{equation}
Thus the decile average is
\begin{equation}
     \langle E \rangle_d = \frac{\sum_{c} \langle E \rangle_c \,n_c}{\sum_{c} n_c}\ ,
  \label{eq: weighted_avg_e}
\end{equation}
where $c$ indexes the fluid cells in a decile bin (indexed by $d$). We plot the decile averages at the median (anti)neutrino fraction within a decile, not the center of the decile, since the (anti)neutrino fractions are not evenly distributed, especially for the lower temperature fluid cells we analyze in Sec.~\ref{sec: compare_mc_fd_warm}.  

The results confirm that at $T\approx 60\,\MeV$ the neutrinos and antineutrinos are
well thermalized: the decile averages agree well with the thermalized-neutrino approximation, and even the individual cell averages show only moderate scatter around that prediction.

\subsubsection{Average Neutrino Absorption Opacity}
\label{sec:hot-nu-avg-opacity}

In the middle row of Fig.~\ref{fig: high_temp_grid}, we plot the average absorption opacity for neutrinos (left panel, pink) and antineutrinos (right panel, purple) as a function of the (anti)neutrino fraction. The pale dots are averages (within the same MC fluid cells as in our average energy comparison above) of the opacity $\langle \kappa \rangle_c^\text{MC}$ computed using Eq.~\eqref{eq: avg_opacity}.
The solid pink/purple lines are the predictions of the thermalized-neutrino approximation for the average (anti)neutrino opacity, obtained from Fermi--Dirac distributions. 
 
We plot decile medians of the MC data for neutrinos (pink diamonds) and antineutrinos (purple diamonds) obtained by taking the median of the distribution of (anti)neutrino average opacities from individual fluid cells $\langle \kappa \rangle_c^\text{MC}$ within a decile, where the deciles are determined by
the (anti)neutrino fraction. We take medians (in contrast to our approach for the average energy) to be consistent with the lower-temperature analysis of Sec.~\ref{sec: compare_mc_fd_warm}, where for the opacity we show medians because the means are
skewed by a few outlier fluid cells. At the high temperature analyzed in this section, the mean and median are similar.

The results are consistent with the expectation that at $T\approx 60\,\MeV$ the neutrinos and antineutrinos are well thermalized: the decile medians agree well with the thermalized-neutrino approximation.
\begin{figure*}[htbp]
\centering

\includegraphics[width=0.329\linewidth]{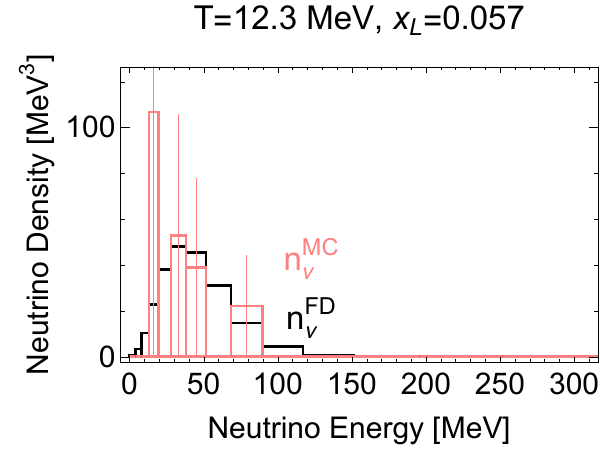}
\includegraphics[width=0.329\linewidth]{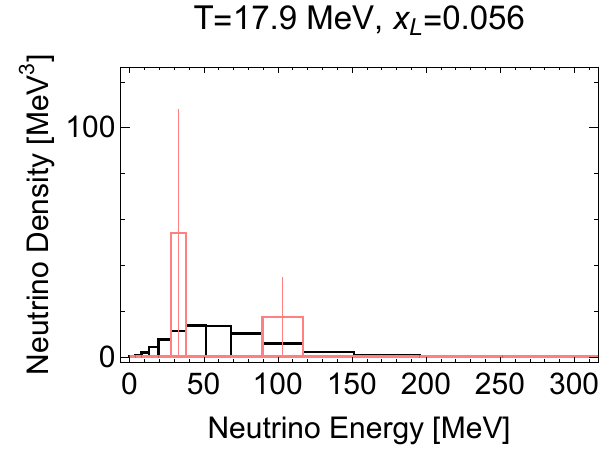}
\includegraphics[width=0.329\linewidth]{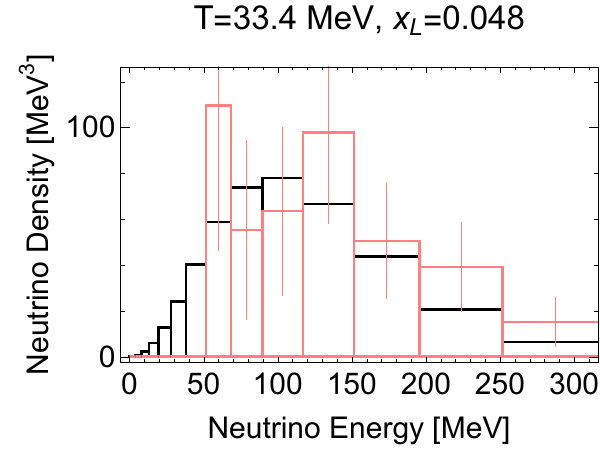}


\includegraphics[width=0.329\linewidth]{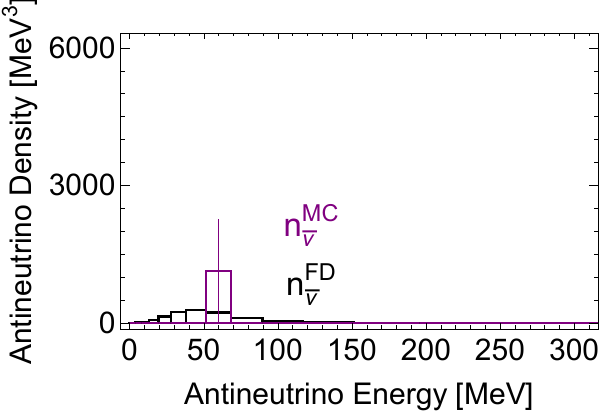}
\includegraphics[width=0.329\linewidth]{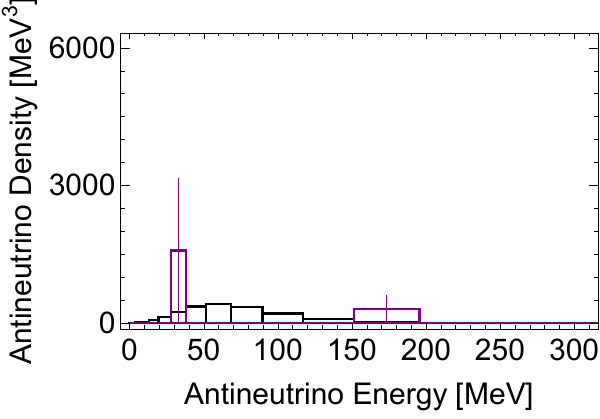}
\includegraphics[width=0.329\linewidth]{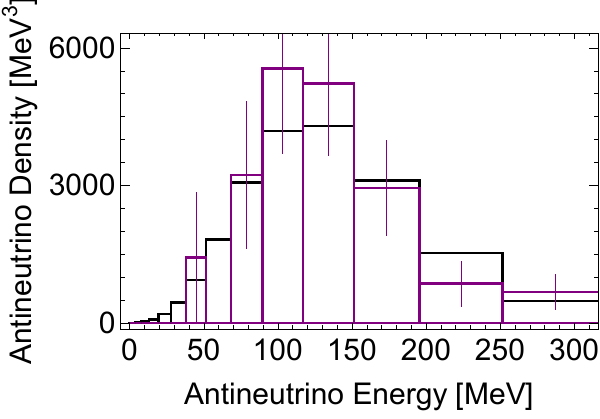}

\caption{
Examples of neutrino (top row) and antineutrino (bottom row) spectra for three selected fluid cells with different temperatures and lepton fractions and with fixed baryon density and proton fraction ($n_B=1.78\,\nsat$, $x_p=0.057$). The Monte Carlo (MC) simulation data are shown as pink/purple histograms. The Fermi--Dirac distributions predicted by the thermalized-neutrino approximation are shown as black histograms. The (anti)neutrino chemical potentials were chosen such that the (anti)neutrino number densities are the same as the MC data. Each column represents data from a single fluid cell.
}

\label{fig: sample_nu_dist}
\end{figure*}
\subsubsection{Net Rate of Neutrino Absorption}
\label{sec:hot-neutrino-absorption}

In the bottom row of Fig.~\ref{fig: high_temp_grid}, we plot the net neutrino absorption rate per baryon $\gamma_\nu = dx_{\nu}/dt$
(left panel, pink) and net antineutrino absorption rate per baryon $\gamma_\nubar = dx_{\nubar}/dt$
(right panel, purple) as a function of the (anti)neutrino 
fraction $x_{\nu,\nubar}$. The pale dots are net rates for individual MC fluid cells computed using Eqs.~\eqref{eq:gamma-nu} and \eqref{eq: summed_ec} or Eqs.~\eqref{eq:gamma-nubar} and \eqref{eq: summed_net_anu}. 
The solid pink/purple lines are the predictions of the thermalized-neutrino approximation for the rates, obtained from Fermi--Dirac distributions, and the blue horizontal lines are the predictions of the free-streaming-neutrino approximation, where there are no neutrinos in initial states and no Pauli blocking of final states. 

We plot decile medians of the MC data for neutrinos (pink diamonds) and antineutrinos (purple diamonds) obtained by taking the median of the distribution of net (anti)neutrino absorption rates from individual fluid cells within a decile, where the deciles are determined by the (anti)neutrino fraction. As with the average absorption opacity in the previous subsection, we plot the decile medians rather than means.

We also include along the upper $x$-axis the (anti)neutrino chemical potentials in the thermalized-neutrino approximation to give an estimate of how far from equilibrium each process in Eq.~\eqref{eq: xp_change_processes} is. Note that the (anti)neutrino chemical potentials do not scale linearly with their number densities.

The thermalized-neutrino approximation (solid pink/purple line) predicts that the cells included in these figures have a net (anti)neutrino absorption rate that rises from negative values to positive values as the (anti)neutrino fraction increases. Equilibrium occurs when $x_\nu = 1.4\times 10^{-3}$ and $x_\nubar = 3.6\times 10^{-2}$, since that is where the predicted net rates are zero. 
We find that, as expected at high temperature, the neutrinos and antineutrinos follow the predictions of the thermalized-neutrino approximation and do not follow the free-streaming-neutrino approximation.

One noticeable feature of the plot is that in many fluid cells the (anti)neutrinos are consistent with the thermalized-neutrino approximation, but deviate from the chemical equilibrium (anti)neutrino fraction. This occurs despite the fact that the net rate of (anti)neutrino absorption per baryon for those cells is very fast: a few to tens of thousands $\ms^{-1}$, corresponding to a chemical equilibration time shorter than a microsecond. Our results are a consequence of the use of MC methods. MC sampling of the distribution function of (anti)neutrinos in regions where (anti)neutrinos are expected to be in thermal equilibrium with the fluid (e.g.,~the $T=62.5\,\MeV$ fluid cells we consider here) are effectively MC samplings (here containing up to $P_c \sim 300$ MC packets per fluid cell) of their thermalized distribution. We therefore expect on the order of $1/\sqrt{P_c} \sim 10\%$ scatter in any given realization of the (anti)neutrino distribution function within a cell. In a cell where a realization of the neutrino distribution function over(under)-samples the number of (anti)neutrinos, the net rate of (anti)neutrino absorption per baryon will then be positive (negative), driving the distribution back towards equilibrium on a time scale comparable to the time step used in the simulation. 
The physical equilibration time scale may be shorter than a numerical time step, but it is limited to the time step scale to guarantee numerical stability of the evolution, see Ref.~\cite{Foucart:2021mcb}. 

Is is noteworthy that, to within the uncertainty scatter of the fluid cells, the (anti)neutrino chemical potentials required to achieve steady state (zero net absorption rate) obey the equilibrium condition $\mu_\nubar=-\mu_\nu$. This is true even though no direct thermalization channels between the neutrinos and antineutrinos were included in the simulation, i.e., no electron (anti)neutrino pair processes and the scattering processes (see Eq.~\eqref{eq: scattering_processes}) were implemented using the elastic approximation. In the elastic approximation, the initial and final energies of the (anti)neutrinos are the same. Chemical equilibrium is instead established individually through the (anti)neutrino absorption and emission processes, shown in Eq.~\ref{eq: xp_change_processes} and the bottom row of Fig.~\ref{fig: high_temp_grid}.


\begin{figure*}[htbp]
\centering

\includegraphics[width=0.329\linewidth]{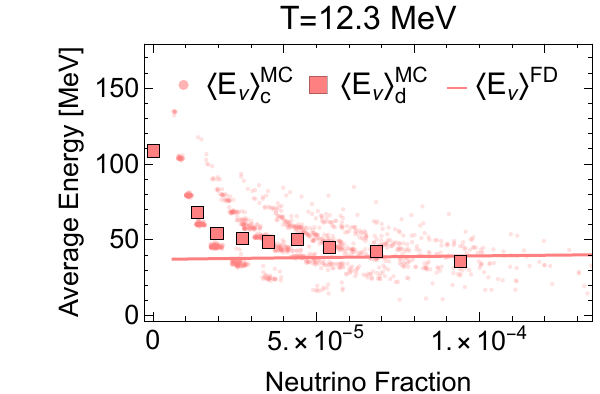}
\hfill
\includegraphics[width=0.329\linewidth]{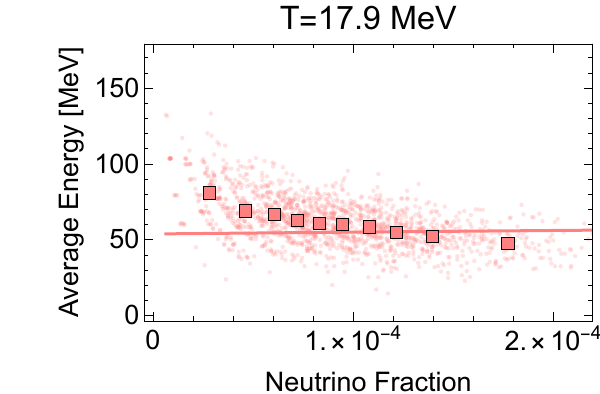}
\hfill
\includegraphics[width=0.329\linewidth]{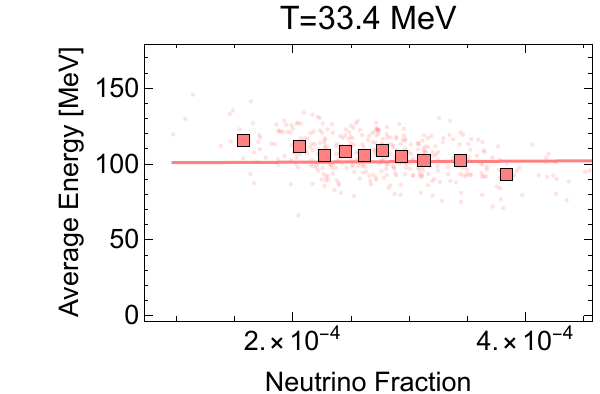}

\vspace{0.5em}

\includegraphics[width=0.329\linewidth]{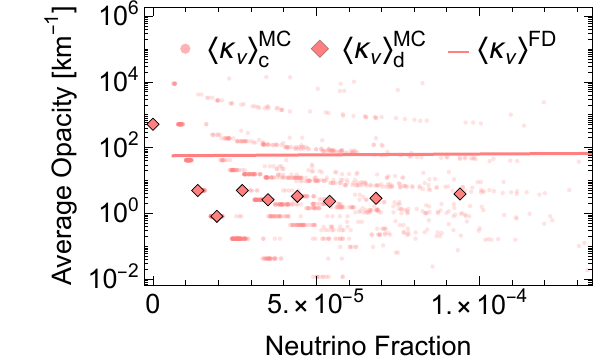}
\hfill
\includegraphics[width=0.329\linewidth]{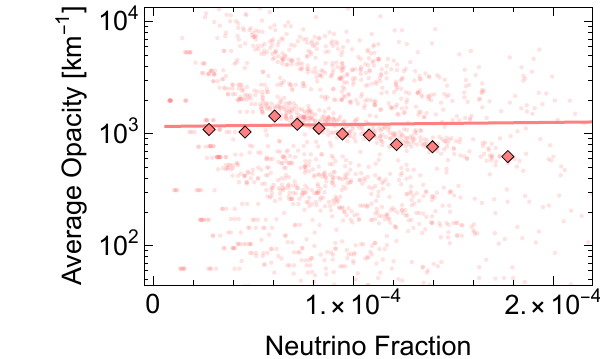}
\hfill
\includegraphics[width=0.329\linewidth]{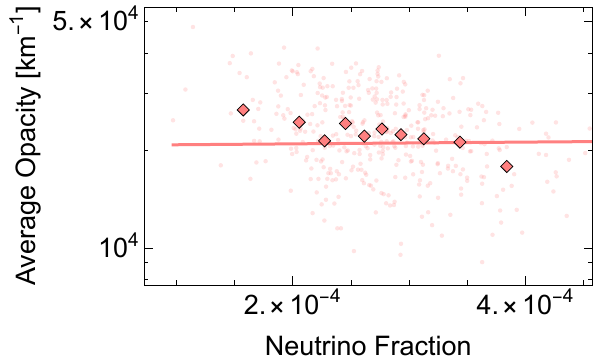}

\vspace{0.5em}

\includegraphics[width=0.329\linewidth]{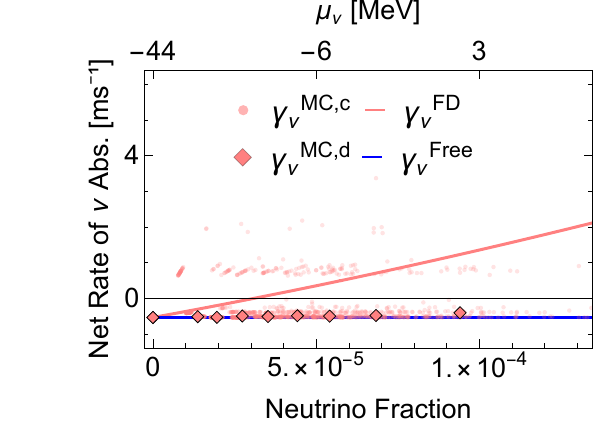}
\hfill
\includegraphics[width=0.329\linewidth]{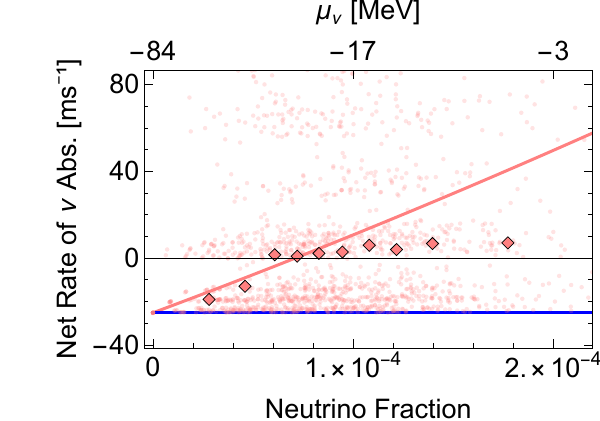}
\hfill
\includegraphics[width=0.329\linewidth]{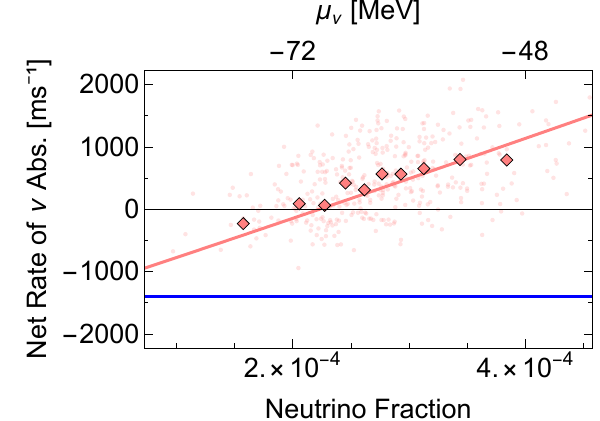}

\caption{Properties of the neutrino population in fluid cells with baryon density $n_B=1.78\,\nsat$ and proton fraction $x_p=0.057$, and with temperatures $T=12.3\,\MeV$ (left column), $T=17.9\,\MeV$ (middle column), and $T=33.4\,\MeV$ (right column), comparing Monte Carlo (MC) simulation data with predictions from thermalized-neutrino (FD) and free-streaming-neutrino (Free) approximations.\\
(a) Top row: Average neutrino energy as a function of neutrino fraction, showing individual MC fluid cells (pale pink dots), averages over cells in neutrino fraction deciles, and the thermalized-neutrino approximation (solid pink line).\\
(b) Middle row: Average opacity, following the same conventions as the top row, except the pink diamonds are decile medians.\\
(c) Bottom row: Net rate of neutrino absorption per baryon as a function of neutrino fraction, following the same conventions as the top row, except the pink diamonds are decile medians. Blue lines are the predictions of the free-streaming-neutrino approximation.
}
\label{fig: nu_moderate_temp_grid}
\end{figure*}

\begin{figure*}[htbp]
\centering

\includegraphics[width=0.329\linewidth]{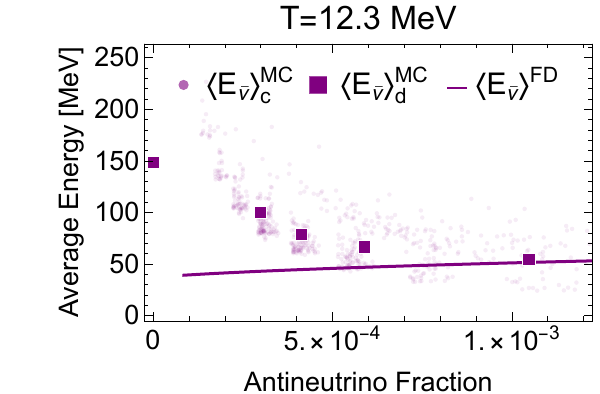}
\includegraphics[width=0.329\linewidth]{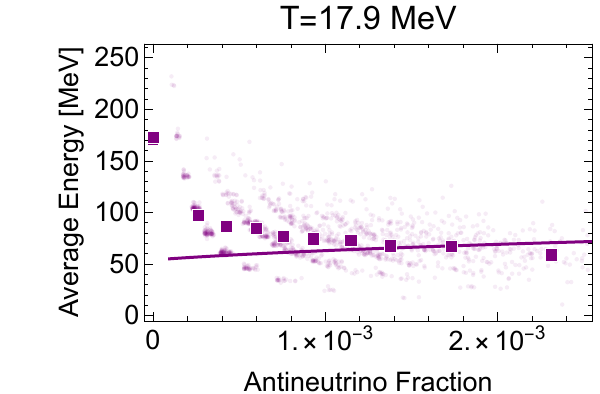}
\includegraphics[width=0.329\linewidth]{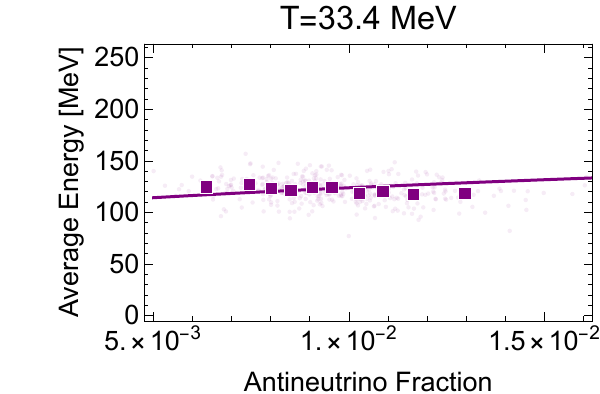}

\vspace{0.5em}

\includegraphics[width=0.329\linewidth]{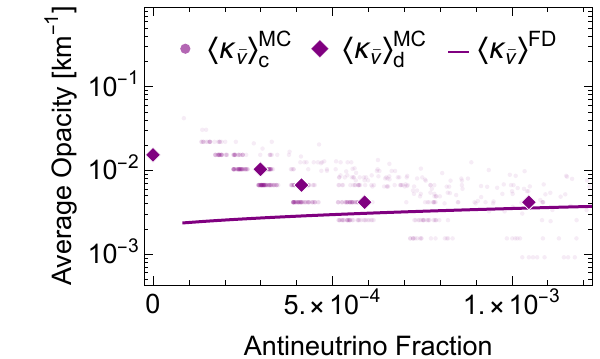}
\includegraphics[width=0.329\linewidth]{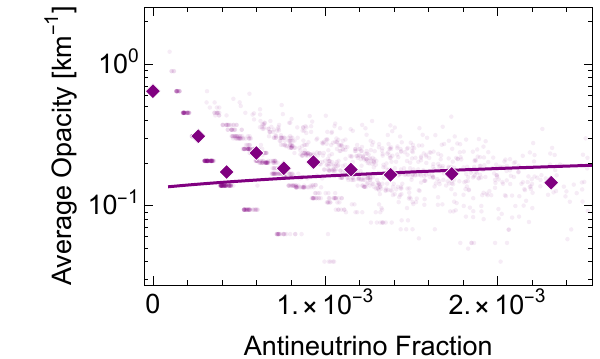}
\includegraphics[width=0.329\linewidth]{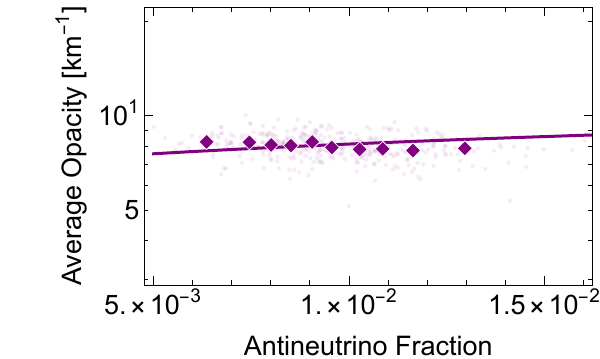}

\vspace{0.5em}

\includegraphics[width=0.329\linewidth]{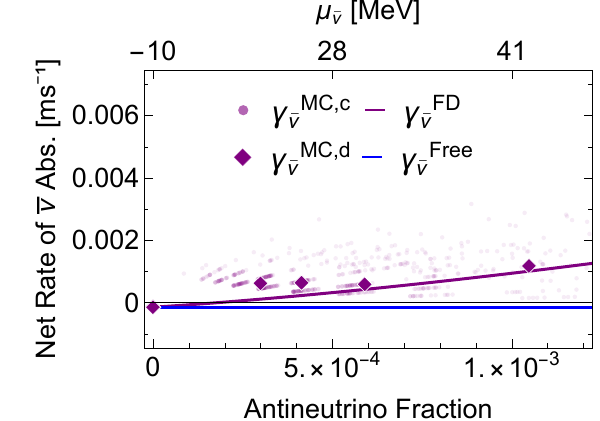}
\includegraphics[width=0.329\linewidth]{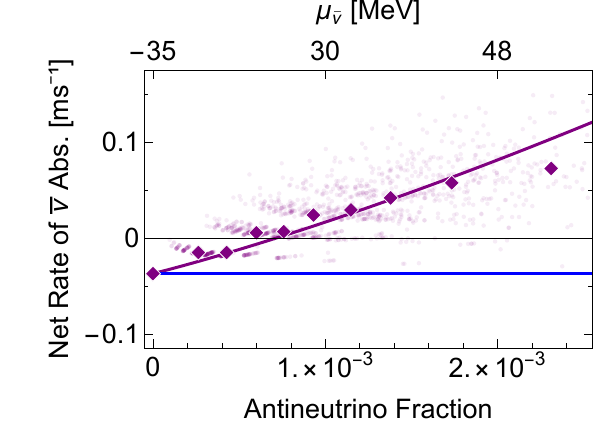}
\includegraphics[width=0.329\linewidth]{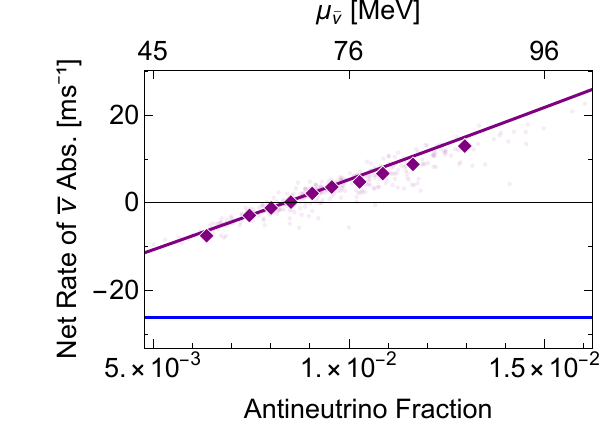}

\caption{Properties of the antineutrino population in fluid cells with baryon density $n_B=1.78\,\nsat$ and proton fraction $x_p=0.057$, and with temperatures $T=12.3\,\MeV$ (left column), $T=17.9\,\MeV$ (middle column), and $T=33.4\,\MeV$ (right column), comparing Monte Carlo (MC) simulation data with predictions from thermalized-neutrino (FD) and free-streaming-neutrino (Free) approximations. See text and the caption of Fig.~\ref{fig: nu_moderate_temp_grid}.
}
\label{fig: nubar_moderate_temp_grid}
\end{figure*}

\subsection{Warm Temperatures}
\label{sec: compare_mc_fd_warm}
We now focus on warm ($T$ between $10$ and $35\,\MeV$) fluid cells, where, as noted in Sec.~\ref{sec: thermalization}, we expect neither the free-streaming nor the thermalized-neutrino approximation will be valid, so the (anti)neutrino spectra and observables may show significant departures from both approximations.

\subsubsection{Warm Neutrino Energy Spectra}
In Fig.~\ref{fig: sample_nu_dist}, we show examples of neutrino and antineutrino energy spectra. 
We compare the MC simulation spectra (pink/purple histograms) with Fermi--Dirac distributions from the thermalized-neutrino approximation (black histograms), discretized into the same energy bins as used in the MC simulation, see Table~\ref{tab: energy_bins}. 

Each column of Fig.~\ref{fig: sample_nu_dist} is from an individual fluid cell. These three cells were chosen to have different temperatures and lepton fractions, but all three cells have the same baryon density and proton fraction, $n_B=1.78\,\nsat$ and $x_p=0.057$, within $\pm\, 5\%$ tolerance. We chose the median lepton fraction for fluid cells that contain neutrinos and antineutrinos at each temperature because no single lepton fraction is represented at all three temperatures. Each fluid cell shown is representative of the (anti)neutrino spectra seen across all fluid cells at each temperature slice. The results are similar for typical fluid cells across the density range $1\,\nsat$ to $3\,\nsat$. 

We note the following features:\\
(a) The number of antineutrinos is much greater than the number of neutrinos. This is expected because in these cells the lepton fraction is smaller than the electron fraction.\\
(b) As the temperature rises, the MC distributions appear to become more consistent with the thermalized-neutrino approximation. At the lowest temperature $T=12.3\,\MeV$, the spectra do not resemble thermal spectra at all: typically, there is only one MC packet in an energy bin, or none at all. At the highest temperature $T=33.4\,\MeV$, the MC (anti)neutrino distribution looks close to thermal equilibrium.

\subsubsection{Temperature Dependence of Observables}
The features noted above in our brief examination of a few representative fluid cells lead us to expect that at temperatures around $10$--$20\,\MeV$ individual fluid cells will show large deviations from the thermalized-neutrino approximation, but the agreement will improve significantly if we look at warmer fluid cells. It is also possible that statistical properties of ensembles of fluid cells will look thermalized at lower temperatures. To test these inferences, we will focus on the same observables as in Sec.~\ref{sec: average_nu_e_hot_results}: (anti)neutrino average energies, average opacities, and net rate of absorption per baryon. We will study how the neutrinos and antineutrinos become more thermalized as the temperature rises from around $10\,\MeV$ to $35\,\MeV$.

Our results are plotted in Fig.~\ref{fig: nu_moderate_temp_grid} (for neutrinos) and Fig.~\ref{fig: nubar_moderate_temp_grid} (for antineutrinos). In both figures, we show results for a subset of the fluid cells: those with baryon density $n_B=1.78\,\nsat$ and proton fraction $x_p=0.057$, within $\pm\,5\%$ tolerance. These particular values of $n_B$ and $x_p$ were chosen because they lie on the \texttt{NuLib} grid we use to evaluate the opacities and rates and allow us to keep the density and proton fraction fixed while varying the temperature in the $10{-}35\,\MeV$ range, see Fig.~\ref{fig: fluid_xp_contours}. We found similar behavior for $x_p=0.045$ and for other densities $n_B=1-3\,\nsat$. The plots follow the same conventions and definitions as for Fig.~\ref{fig: high_temp_grid}, see the discussion in Sec.~\ref{sec: average_nu_e_hot_results}.

In each figure, each row contains results for one observable: average energy (top row), average opacity (middle row), and net absorption rate per baryon (bottom row), all plotted as a function of the (anti)neutrino fraction in the fluid cells. The columns track the temperature dependence: the first column is for cells with temperature $T=12.3\,\MeV$, the second column for $T=17.9\,\MeV$, and the third column for $T=33.4\,\MeV$.

The main conclusion from Figs.~\ref{fig: nu_moderate_temp_grid} and \ref{fig: nubar_moderate_temp_grid} is that as the temperature rises to $33.4\,\MeV$, the neutrinos and antineutrinos approach the thermalized-neutrino approximation. We expect this behavior because both the neutrino and antineutrino mean net displacement decrease as the temperature increases, see Fig.~\ref{fig:mfp-diffusion}. 

The fluid cells are split into deciles based on their (anti)neutrino fractions. In some panels of Figs.~\ref{fig: nu_moderate_temp_grid} and \ref{fig: nubar_moderate_temp_grid}, not all ten decile mean/median symbols may be visible. Specifically, in
the first column ($T=12.3\,\MeV$) in Fig.~\ref{fig: nu_moderate_temp_grid}, the first neutrino fraction decile is absent because it consists of fluid cells that contain no neutrinos. In the first column ($T=12.3\,\MeV$) of Fig.~\ref{fig: nubar_moderate_temp_grid}, the first five antineutrino deciles are absent because they consist of fluid cells that contain no antineutrinos. 
This is not surprising, since the energy density in antineutrinos in these cells is on the order of the energy density of a single MC packet.

As mentioned in Sec.~\ref{sec: compare_mc_fd_hot}, for the average (anti)neutrino energy we plot decile means, whereas for the average opacity and net absorption rate per baryon we plot decile medians. This is because at $T=17.9\,\MeV$ and even more strongly at $T=12.3\,\MeV$ the distribution of average opacities and net absorption rates in a decile have a tail of a few fluid cells that have much higher values, which skew the mean to a much higher value than the median.

We find that for fluid cells in Fig.~\ref{fig: nu_moderate_temp_grid}, with $T=12.3\,\MeV$ and $T=17.9\,\MeV$, the average opacity and net absorption rate per baryon do not look comparable to the thermalized-neutrino approximation, even though the average energy of the neutrinos looks comparable to the thermalized-neutrino approximation. This is possible because the average opacity and the net absorption rate are more sensitive to the shape of the neutrino distribution than the average energy.

In each observable we show, there are some striations and groupings due to the finite number of packets used by the MC scheme within any given cell. While the cells at $T=62.5\,{\rm MeV}$ (discussed in Fig.~\ref{fig: high_temp_grid}) contain $\sim 300$ packets per cell, at the lower temperatures we may only have a few packets per cell. 
In the simulation, packets have varying numbers of (anti)neutrinos, but each packet has approximately the same fluid-frame energy $E_\text{packet}$. It follows from Eq.~\eqref{eq: avg_e_using_e_tot} that a cell with $P_{c}$ packets and (anti)neutrino fraction $x_{\nu,\nubar}$ will have an average (anti)neutrino energy 
\begin{equation}
 \langle E_{\nu,\nubar}\rangle_c = \dfrac{P_c E_\text{packet}}{N_B\, x_{\nu,\nubar}}\,,
\end{equation}
where $N_B$ is the number of baryons in cell $c$. This explains the hyperbola-shaped striations seen in the average energy at low (anti)neutrino fraction in some of the lower-temperature panels: they are numerical artifacts arising when most of the cells with a given (anti)neutrino fraction have a small number of packets.
For example, in Fig.~\ref{fig: nu_moderate_temp_grid} at $12.3\,{\rm MeV}$, the leftmost hyperbolic band corresponds to $P_{\rm c}=1$, and subsequent bands are seen as $P_{\rm c}$ increases from left to right. These features become less visible at higher temperatures, when most cells have a larger number of packets and a substantial (anti)neutrino fraction. The deviation of the average energy from the expected Fermi--Dirac value at low (anti)neutrino fraction is thus not necessarily a sign that the MC (anti)neutrinos are out of thermal equilibrium. This is instead because the lowest non-zero neutrino fractions are naturally obtained in cells with a single packet representing high-energy neutrinos. 
In the average opacity, the groupings are still present, but are now spaced farther apart. This is because the (anti)neutrino opacity changes dramatically over a small range of energies, see Fig.~\ref{fig:mfp-diffusion}. 

In the bottom row of Figs.~\ref{fig: nu_moderate_temp_grid} and \ref{fig: nubar_moderate_temp_grid}, the diagonal lines correspond to fluid cells that have all (anti)neutrinos in a single energy bin, regardless of the number of packets. These groupings are more prominent at lower temperatures, where there are fewer MC packets in each fluid cell.

\section{Conclusions}
\label{sec:conclusions}
We compared neutrino distributions derived from a neutron star merger simulation $1\,\text{ms}$ after merger, using a Monte Carlo (MC) neutrino transport scheme, with neutrino population assumptions that are commonly used: the free-streaming-neutrino and the thermalized-neutrino approximations. To make this comparison, we computed the (anti)neutrino average energy, average absorption opacity, and net rate of absorption per baryon. 

We found that at $T\approx 60\,\MeV$, which is the highest temperature reached by substantial regions in the MC simulation data, our observables computed using MC (anti)neutrino distributions agreed well with the thermalized-neutrino approximation, see Fig.~\ref{fig: high_temp_grid}.

We then investigated the range of moderate temperatures ($T \approx 10-35\,\MeV$) in the MC simulation data.
We found that as the temperature increases through this range, the agreement between the MC data and the thermalized-neutrino approximation improves, see Figs.~\ref{fig: nu_moderate_temp_grid} and \ref{fig: nubar_moderate_temp_grid}.
At the lower end ($T=12.3\,\MeV$), the net absorption rate per baryon for neutrinos agrees well with the free-streaming-neutrino approximation (and partial agreement for the antineutrinos). At the higher end ($T=33.4\,\MeV$), all of our observables show good agreement with the thermalized-neutrino approximation, at least when considering means or medians over many fluid cells. Individual fluid cells show considerable scatter around the predicted values.

The calculations reported here represent an early stage in the project of capturing neutrino transport in mergers via MC simulations. There are various limitations that we hope will be addressed in future work.

Firstly, we analyzed one time slice ($t=1\,\ms$ after merger). It would be useful to explore later time slices to see how the (anti)neutrino distributions evolve and to compare with the results of Ref.~\cite{Espino:2023dei}. 

Secondly, the simulation shows some numerical artifacts, such as the striations seen in Figs.~\ref{fig: nu_moderate_temp_grid} and \ref{fig: nubar_moderate_temp_grid}. These will diminish as
MC simulations use more packets or develop improved methods for tracking the neutrino physics. An example would be the use of weights for the various MC packets that are optimized for the observable that we want to measure, see, e.g., Ref.~\cite{Foucart:2025nub} for a more in-depth discussion.

Finally, the neutrino-matter interaction library used in the merger simulation we considered, \texttt{NuLib}~\cite{nulib}, makes various assumptions that are not valid for neutron star mergers. 
It would be valuable to perform further
analysis using a more up-to-date neutrino-matter interaction library, e.g., Refs.~\cite{Ng:2023syk, Chiesa:2024lnu}, that includes reactions that are important for creating antineutrinos at low temperature and reactions that can directly equilibrate neutrinos and antineutrinos.

Our findings are especially relevant to simulations that use schemes where a thermalized form of the (anti)neutrino distribution must be assumed, see Sec.~\ref{sec:intro}. 
Our results provide a tentative indication that the assumption of a thermalized form for the neutrino distribution may be unjustified at temperatures below about $30\,\MeV$.
\section{Acknowledgments}
We thank Leonardo Chiesa, Peter Hammond, Steven Harris, Marco Hofmann, Harry Ng, and Ziyuan Zhang for useful discussions.
MGA and LB are partly supported by the U.S. Department of Energy, Office of Science, Office of Nuclear Physics, under Award No.~\#DE-FG02-05ER41375. 
A.H.~acknowledges financial support by the UKRI under the Horizon Europe Guarantee project EP/Z000939/1.
F.F. is supported by the Department of Energy, Office of Science, Office of Nuclear
Physics, under contract number DE-SC0020435 and by
NASA through grant 80NSSC22K0719.

\bibliographystyle{JHEP}
\bibliography{reflist}

\end{document}